\documentclass[%
 aip,
 cha,%
 amsmath,amssymb,
reprint,%
]{revtex4}

\usepackage{graphicx}
\usepackage{dcolumn}
\usepackage{bm}
\begin{document}


\title{Desynchronization bifurcation of coupled nonlinear dynamical systems}

\author{Suman Acharyya}
\email{suman@prl.res.in}
\author{R. E. Amritkar}%
\email{amritkar@prl.res.in}
\affiliation{%
Theoretical Physics Division, 
Physical Research Laboratory\\
Navrangpura, Ahmedabad-380009
}%

\begin{abstract}
We analyze the desynchronization bifurcation in the coupled R\"ossler oscillators. 
After the bifurcation the coupled oscillators move away from each other with a square root dependence on the parameter. 
We define system transverse Lyapunov exponents and in the desynchronized state one is positive while 
the other is negative implying that one oscillator is trying to fly away while the other is holding it.
We give a simple model of coupled integrable systems that shows a similar phenomena and can be treated as 
the normal form for the desynchronization bifurcation. We conclude that the desynchronization is a pitchfork bifurcation of the transverse manifold.
\end{abstract}

                             

\maketitle

\begin{quotation}
In this paper we study  desynchronization bifurcation for two chaotic R\"ossler oscillators which are mutually coupled via $x$-component. 
They show an interesting behaviour with increasing coupling strength ($\varepsilon$). 
When coupling strength ($\varepsilon$) is increased beyond some first critical coupling ($\varepsilon_{c1}$), 
the oscillators synchronize. 
As we keep on increasing coupling strength, these oscillators will remain synchronized for some time ($\varepsilon_{c1}<\varepsilon<\varepsilon_{c2}$).
When coupling strength exceeds a second critical value ($\varepsilon_{c2}$) the oscillators desynchronize. 
At this point the largest transverse Lyapunov exponent (TLE) become positive.
To understand this phenomenon in more details, we define systems' transverse Lyapunov exponents (STLE) which are specific to each system.
In the synchronized state the STLE and TLE have similar value and all are negative. 
But in the desynchronized state one of the STLEs is positive and another is negative which implies that 
the perturbation grows about one system while it dies out about the other system, 
i.e. one system is trying to fly away while the other is holding it. 
We present a simple integrable general model with quadratic nonlinearity which shows similar phenomena and 
the nature of this desynchronization can be explored in more details with the help of this model. 
This model can be regarded as a normal form for the desynchronization bifurcation in coupled R\"ossler systems. We also study the cubic nonlinearity and find that in this case both SLTEs are negative.

\end{quotation}

\section{INTRODUCTION}

Chaotic systems when coupled in some fashion or driven by same external signal, 
synchronize as the coupling strength increases, 
and are said to be identically or completely synchronized when 
the variables of the systems become equal~\cite{PTP.69.32,PTP.70.1240,Sov.Tech.Phys.Lett.15,Izv.Vys.Uche.Zv.Rad.29,PhysRevLett.64.821}. 
We can observe other types of synchronization, such as phase synchronization~\cite{PhysRevLett.76.1804,PhysRevLett.80.1642}, 
lag synchronization~\cite{PhysRevLett.78.4193}, generalized synchronization~\cite{PhysRevE.51.980,PhysRevLett.76.1816} etc. 
Here, we restrict ourselves to identical synchronization, though some of the results may have more general validity. 

The condition for identical synchronization, or for simplicity synchronization, can be obtained by linear stability analysis. 
The phase space of the coupled system can be split into two manifolds, the synchronization manifold and the transverse manifold. 
The synchronization takes place when the transverse Lyapunov exponents (TLEs) become negative \cite{PhysRevLett.74.4185,PhysRevE.58.347}.

An interesting situation arises when two R\"ossler oscillators are coupled with each other. 
There are two critical coupling constants, $\varepsilon_{c1}$, and $\varepsilon_{c2}$. 
For $\varepsilon < \varepsilon_{c1}$, the oscillators are desynchronized. 
They are synchronized for $\varepsilon_{c1} < \varepsilon <\varepsilon_{c2}$ and are again desynchronized for $\varepsilon > \varepsilon_{c2}$. 
In the range $\varepsilon_{c1} < \varepsilon <\varepsilon_{c2}$, all the TLEs are negative. 
While outside this range one of the TLEs is positive \cite{PhysRevLett.74.4185}. 
Similar considerations apply when one considers a system of coupled R\"ossler systems on a network. 
When one couples several identical chaotic systems in an array, 
the desynchronization bifurcation at $\varepsilon_{c2}$ can be identified as a short wavelength bifurcation, 
where the shortest spatial wavelength mode becomes unstable \cite{PhysRevLett.74.4185,PhysRevE.58.347}.

The purpose of the present paper is to study and understand the desynchronization bifurcation. 
We find that for $\varepsilon > \varepsilon_{c2}$, 
the attractors of the two coupled systems split and start drifting away from each other and 
the rate of drift is proportional to $\sqrt{\varepsilon - \varepsilon_{c2}}$. 
We introduce system transverse Lyapunov exponents (STLEs) and we find that  
the largest STLE for one system becomes positive while that for the other system becomes negative. 
It implies that when desynchronization bifurcation takes place the perturbation transverse to the synchronization manifold grows about one attractor
while it dies out about the other attractor.  
Next, we construct a simple model of coupled integrable systems which obeys similar properties. 
We are able to analytically derive the properties of the desynchronization bifurcation in our model. 
The model can be considered to be the normal form for the desynchronization bifurcation. In the model we study both quadratic and cubic nonlinearites and we find that STLEs are useful to distinguish between the two nonlinearities. The quadratic nonlinerity gives the results corresponding to the desynchronization bifurcation in R\"ossler system.
Form this study we identify the desynchronization bifurcation in coupled R\"ossler systems as a pitchfork bifurcation of the transverse manifold. 
The present paper is divided in the following sections.  In section~\ref{stability-analysis} we study linear stability analysis  
of two $n$ dimensional systems. In section~\ref{stle-intro} we define 
systems' transverse Lyapunov exponents and develop an algorithm to calculate STLE. 
In section~\ref{numerical-expt} we presented numerical results on R\"ossler oscillators. 
We proposed simple integrable model in section~\ref{models} which shows similar behaviour and 
show how this model can be treated as a normal form of this desynchronization bifurcation. 

\section{Desynchronization bifurcation} 

We first consider the linear stability analysis of the synchronized state of two coupled dynamical systems. Next, we introduce system transverse Lyapunov exponents. Then, these are used to study the desynchronization bifurcation in the coupled R\"ossler systems.

\subsection{\label{stability-analysis}Linear stability analysis of synchronized state of two coupled dynamical systems}

Consider an $n$-dimensional autonomous dynamical system,
\begin{equation}
\dot{x} = f(x), \label{eq:one}
\end{equation}
and couple this system with an identical dynamical system $y$,
\begin{eqnarray}
\dot{x} & = & f(x) + \varepsilon_{1}\Gamma(y - x) \nonumber \\
\dot{y} & = & f(y) + \varepsilon_{2}\Gamma(x - y) \label{eq:two}
\end{eqnarray}
where, $\varepsilon_{1}$ and $\varepsilon_{2}$  are scalar coupling parameters. $\Gamma$ is known as the diffusive coupling matrix. 
In general, $\Gamma = \textrm{diag}(\gamma_0,\gamma_1,..., \gamma_{n-1})$,
and defines the components of $x$ and $y$ which are coupled.
The synchronization manifold is defined by $x=y=s$, where $s$ satisfies Eq.~(\ref{eq:one}). Let, $\xi_{x}$ and $\xi_{y}$ be the deviation of $x$ and $y$ from the synchronized solution $s$. We have
\begin{eqnarray}
\dot{\xi}_{x} &=& \nabla f(s) \xi_{x} + \varepsilon_1 \Gamma (\xi_{y} - \xi_{x}) \nonumber \\
\dot{\xi}_{y} &=& \nabla f(s) \xi_{y} + \varepsilon_2 \Gamma (\xi_{x} - \xi_{y}) \label{lsa03}
\end{eqnarray}
These two equations can be also be written as \cite{rang},
\begin{equation}
\dot{{ \xi}} = \nabla f(s) { \xi} + \Gamma  { \xi} G^{T} \label{lsa04}
\end{equation}
where, ${\xi}=(\xi_{x},\xi_{y})$ and $G$ is the coupling matrix. In this case 
\[{ G} = \bigl( \begin{smallmatrix} -\varepsilon_1 & \varepsilon_1 \\ \varepsilon_2 & -\varepsilon_2 \end{smallmatrix} \bigr).\] 
Let, $P_{k}$ be an eigenvector of $G^{T}$ with eigenvalue $\mu_{k}$; ${G^{T}}P_{k}=\mu_{k}P_{k}$. Operating Eq.~(\ref{lsa04}) on $\mu_k$ and defining $\zeta_k = \xi P_{k}$ and we can write an equation for $\zeta_{k}$ as \cite{rang},
\begin{eqnarray}
\dot{\zeta}_{k} = \left[ \nabla f(s) + \varepsilon \mu_{k} \Gamma \right] \zeta_{k}. \label{lsa05}
\end{eqnarray}
Here, the matrix $G$ has two eigenvalues $\mu_{0} = 0$ and $\mu_{1}=-(\varepsilon_1 + \varepsilon_2)$. Thus, Eq.~(\ref{lsa05}) gives the two equations,
\begin{eqnarray}
\dot{\zeta}_{0} &=& \nabla f(s) \zeta_{0} \label{lsa06} \\
\dot{\zeta}_{1} &=& \left[ \nabla f(s) -\mu_1 \Gamma \right]\zeta_{1}. \label{lsa07} 
\end{eqnarray}
Here Eqs.~(\ref{lsa06}) and~(\ref{lsa07}) define motion of small perturbations on the synchronization and transverse manifolds respectively and these can be used to obtain the Lyapunov exponents for the two manifolds. 
The synchronized state will be stable when all the transverse perturbations die with time, i.e. when all the transverse Lyapunov exponents are negative.

\subsection{\label{stle-intro}System's Transverse Lyapunov Exponents}

We now introduce transverse Lyapunov exponents which are specific to the invidual systems $x$ and $y$.

The dynamics of the difference vector
$z = x - y$, is
\begin{eqnarray}
\dot{z} = f(x) - f(y) - (\varepsilon_{1} + \varepsilon_{2})\Gamma z \label{eq:four}
\end{eqnarray}
In Eq.~(\ref{eq:four}) we can expand $f(y)$ in Taylor's series about the co-ordinate of the first system $x$ or $f(x)$ about the second system $y$. This gives us the following two equations.
\begin{eqnarray}
\dot{z} & = & (z \cdot \nabla) f(x) - (\varepsilon_{1} + \varepsilon_{2})\Gamma z \label{stle-x} \\
\label{stle}
\dot{z} & = & (z \cdot \nabla) f(y) - (\varepsilon_{1} + \varepsilon_{2})\Gamma z \label{stle-y}
\end{eqnarray}
where we neglect the higher order terms.
In the synchronized state, Eqs.~(\ref{stle-x}) and~(\ref{stle-y}) are identical and give the transverse Lyapunov exponents.
In the desynchronized state, Eqs.~(\ref{stle-x}) and~(\ref{stle-y}) in general give different exponents and 
we refer to them as system transverse Lyapunov exponents (STLEs) since they are specific to each system and 
denote the largest of them as $\lambda_x$ and $\lambda_y$ respectively. 
For the synchronized state $\lambda_x=\lambda_y$ and they are negative. 
For the desynchronized state $\lambda_x$ may not be equal to $\lambda_y$ and tell us about 
how the difference vector $z$ behaves in the neighborhood of the two systems. 
Note that for the synchronized state these STLEs belong to the actual spectrum of 
Lyapunov exponents of the coupled system, but not for the desynchronized state.

\subsection{\label{numerical-expt}Two coupled R\"ossler systems}

We now take the specific example of two coupled R\"ossler oscillators \cite{Rössler1976397}. 
Denoting the variables of the two systems by $x$ and $y$ the coupled equations are
\begin{eqnarray}
\dot{x}_1 &=& -x_2 - x_3 + \varepsilon (y_1 - x_1) \nonumber \\
\dot{x}_2 &=&  x_1 + a_rx_2 \label{eq:seven}  \\
\dot{x}_3 &=&  b_r + x_3 (x_1 - c_r) \nonumber 
\end{eqnarray}
and a similar set of equations for the other system $y$.
Here, we have coupled only the first component, i.e. $\Gamma = \textrm{diag}(1,0,0)$ and 
we take symmetric coupling, $\varepsilon = \varepsilon_1 = \varepsilon_2$.

\begin{figure}[h]
\includegraphics[width=0.9\columnwidth]{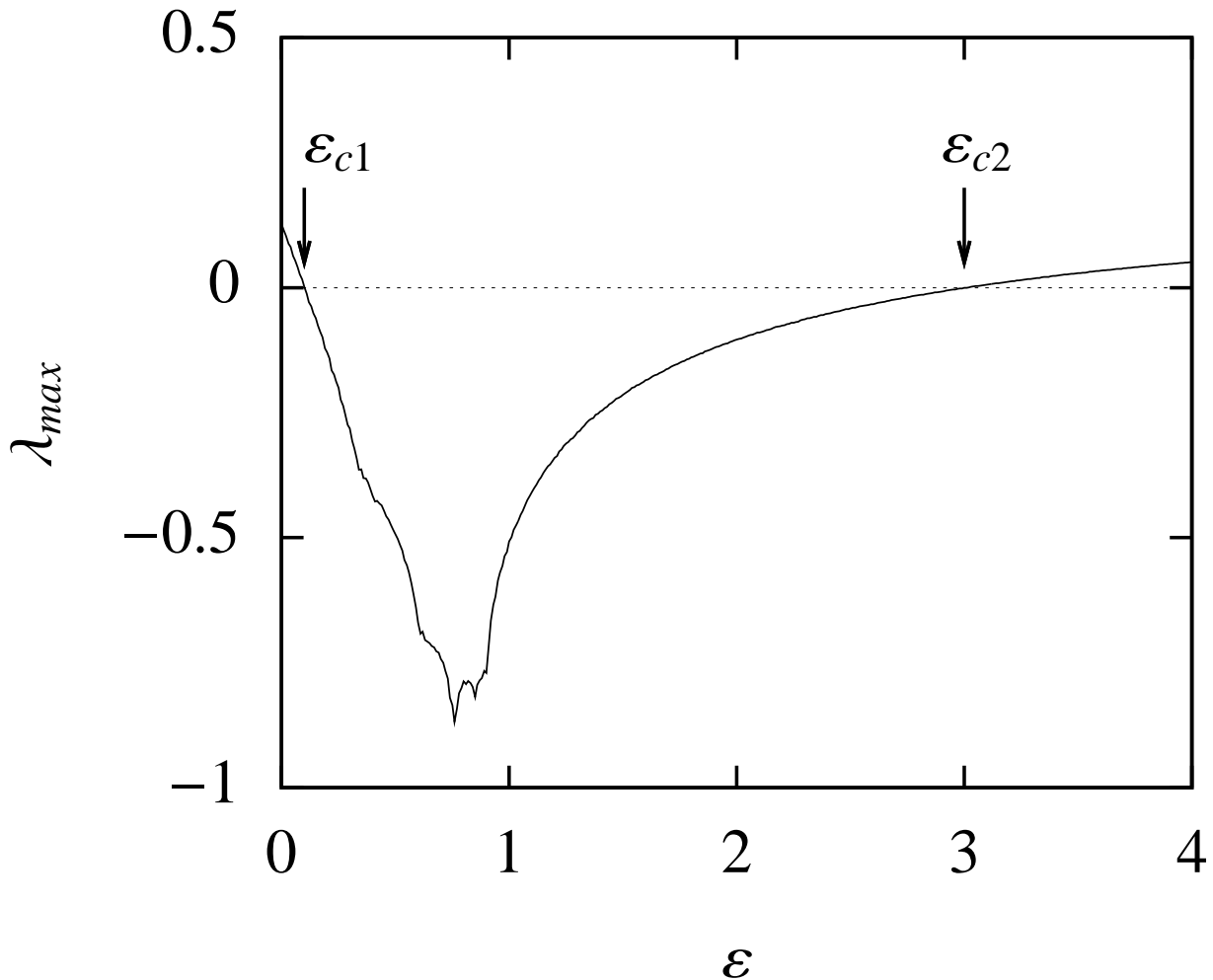}
\caption{\label{fig:msf} The largest transverse Lyapunov exponents, $\lambda_{max}$, of the two coupled chaotic 
identical R\"{o}ssler oscillators are plotted with the coupling parameter $\varepsilon$.
There are two critical couplings $\varepsilon_{c1}(\sim 0.1)$ and $\varepsilon_{c2}(\sim 3.0)$.
In the range $\varepsilon_{c1}<\varepsilon<\varepsilon_{c2}$ the synchronized state is stable.
The desynchronization bifurcation takes place when $\varepsilon>\varepsilon_{c2}$. 
The $\lambda_{max}$ is calculated from Eq.~\ref{lsa07}.
R\"ossler parameters are $a_r = 0.15, \, b_r = 0.2$ and $c_r = 10.0$.
Note that for very large couplings the coupled system become unstable.}
\end{figure}

Fig.~\ref{fig:msf} shows the variation of the largest transverse Lyapunov exponent ($\lambda_{max}$) with coupling strength for 
two mutually coupled identical R\"{o}ssler oscillators. As discussed in the introduction there are two critical coupling constants $\varepsilon_{c1}$ and $\varepsilon_{c2}$.
The synchronized state is stable when $\varepsilon_{c1}<\varepsilon<\varepsilon_{c2}$.
At $\varepsilon = \varepsilon_{c2}$ the system undergo a desynchronization bifurcation.
As we see in Fig~\ref{fig:msf}, $\lambda_{max}$ is positive when $\varepsilon>\varepsilon_{c2}$, which implies the synchronous state is unstable.
To understand this phenomena in details we calculate the systems' transverse Lyapunov exponents ($\lambda_{x}$ and $\lambda_{y}$) introduced in the previous subsection.

\begin{figure}[h]
\includegraphics[width = 0.9\columnwidth]{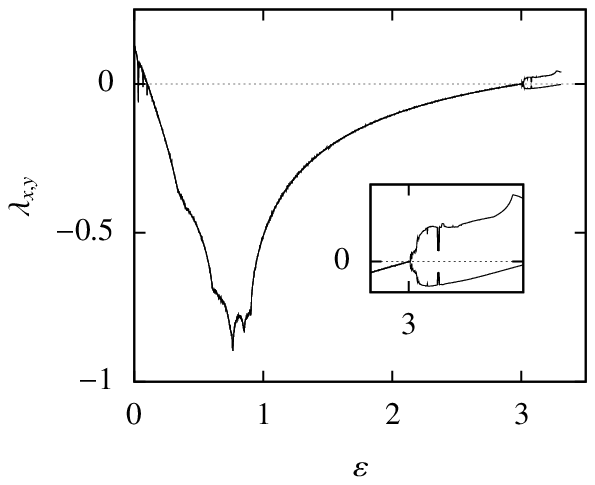}
\caption{\label{fig:one} The largest system transverse Lyapunov exponents, $\lambda_{x,y}$, of the two coupled chaotic 
identical R\"{o}ssler oscillators are plotted with the coupling parameter $\varepsilon$.
The desynchronization bifurcation is observed for large coupling (here at $\varepsilon\sim 3$). 
The inset shows a blowup of $\lambda_x$ and $\lambda_y$ just after the desynchronization takes place.
R\"ossler parameters are $a_r = 0.15, \, b_r = 0.2$ and $c_r = 10.0$.
}
\end{figure}

In Fig.~\ref{fig:one}, the two largest system transverse Lyapunov exponents,
$\lambda_x$ and $\lambda_y$ are plotted as a function of the coupling strength $\varepsilon$. 
As noted before, there are two critical coupling constants, $\varepsilon_{c1}$ and $\varepsilon_{c2}$. 
At both the critical points $\lambda_x = \lambda_y = 0$. For $0 < \varepsilon < \varepsilon_{c1}$, 
the coupled oscillators are desynchronized. The attractors of the two systems overlap and are similar in nature. 
In this region, $\lambda_x \simeq \lambda_y$ and both are mostly positive. 
For, $\varepsilon_{c1} < \varepsilon < \varepsilon_{c2}$, the two R\"ossler oscillators are synchronized. 
Here, $\lambda_x = \lambda_y$ and both are negative. For $\varepsilon > \varepsilon_{c2}$, 
the oscillators become desynchronized. Here, the largest STLEs show an interesting behavior. 
One of STLE becomes positive but the other becomes negative. 
Note that for very large values of $\varepsilon$ the coupled system becomes unstable. 

\begin{figure}[t]
\includegraphics[width = 0.9\columnwidth]{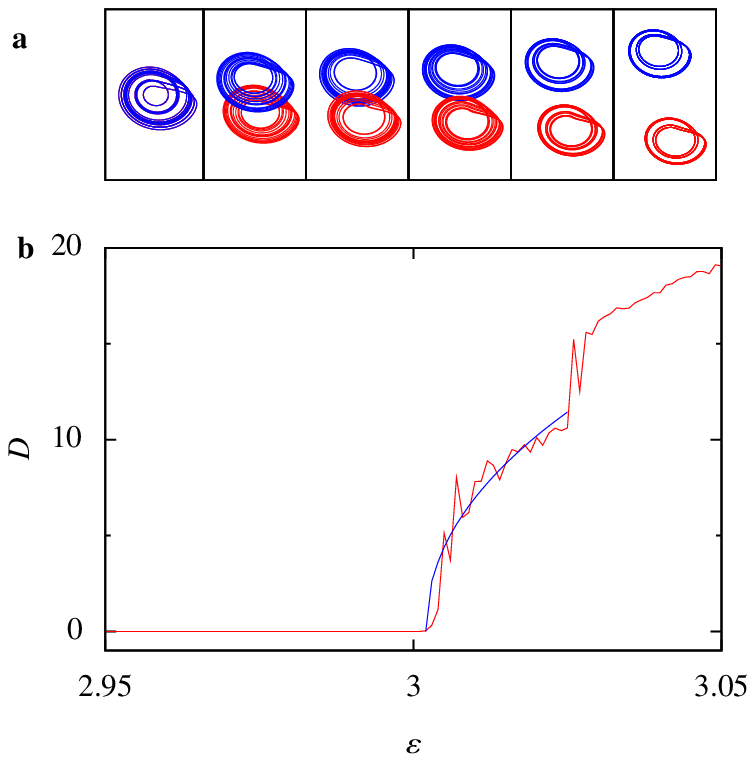}
\caption{\label{fig:four} {\bf a}. 
The projection of the attractors of the two R\"{o}ssler oscillators (red and blue online) on x-y plane for 
$\varepsilon= 3.000, 3.025, 3.050, 3.100, 3.150, 3.200$. 
Note that Eqs.~(\ref{eq:seven}) obey the $x \Leftrightarrow y$ exchange symmetry and hence the attractor obtained by the exchange $x \Leftrightarrow y$, is also a solution. 
{\bf b}. 
The distance $D$, between the centers of the attractors of the 
two R\"ossler oscillators is plotted as a function of the coupling strength $\varepsilon$. 
The continuous curve (blue online) is a power law fit (Eq.~(\ref{eq:eight})) with the exponent $\nu=0.5$. Note that the diastance $D$ is also proportional to the two solutions obtained by the $x \Leftrightarrow y$ exchange symmetry.}
\end{figure}

To understand the result that one STLE is positive and the other is negative, 
let us first look at the phase space plots of the attractors of the two coupled oscillators in Fig.~\ref{fig:four}a. 
The two attractors are identical and overlap at $\varepsilon = \varepsilon_{c2}$. 
As $\varepsilon$ increases the two attractors split and start moving away from each other as shown in Fig.~\ref{fig:four}a. 
Figure~\ref{fig:four}b shows the distance $D$, between the centers of the two attractors as a function of $\varepsilon$. 
For $\varepsilon  > \varepsilon_{c2}$, the distance $D$ shows a power law behavior,
\begin{equation}
D = \gamma (\varepsilon - \varepsilon_{c2})^\nu, \label{eq:eight}
\end{equation}
The fit is shown in Fig.~\ref{fig:four}b and the exponent is $\nu =0.474 \pm 0.054 \sim 0.5$ and the other parameters are
$\gamma = 67.38 \pm 15.75, \varepsilon_{c2} = 3.002 \pm 0.000004$.
The power law behavior is a characteristic feature of a second order phase transition. 

Let us now come back to the result of Fig.~\ref{fig:one}, 
that for $\varepsilon  > \varepsilon_{c2}$ one of the STLE is positive and the other is negative.
These STLEs tell us about the behavior of the distance between the attractors as viewed from each of them in the linear approximation.
Thus, we can say that in the linear approximation one of the attractors is trying to fly away while the other one is trying to hold them together. 
The stability of the coupled system implies that the negative STLE wins the battle. 
It appears that as $\varepsilon$ increases, 
the hold of the negative STLE decreases and hence the two attractors start drifting away from each other 
and for large values of $\varepsilon$ the system becomes unstable.

\section{\label{models}MODEL SYSTEM}

Since R\"ossler oscillators are chaotic it is not easy to decipher the behavior of the desynchronization bifurcation. 
Hence, we now propose a simple model of coupled integrable systems showing a similar desynchronization bifurcation. 
It is easy to see that with one dimensional systems we do not get the desynchronization bifurcation.
Hence the minimum dimension is two.
The proposed model is
\begin{eqnarray}
\dot{ x_1} &=& ax_1 + bx_2 + \varepsilon(y_1 - x_1) \nonumber \\
\dot{ x_2} &=& cx_1 + dx_2 + g(x_1,x_2) \nonumber \\
\dot{ y_1} &=& ay_1 + by_2 + \varepsilon(x_1 - y_1) \nonumber \\
\dot{ y_2} &=& cy_1 + dy_2 + g(y_1,y_2), 
\label{model}
\end{eqnarray}
Here, $a,b,c,d,\alpha,\beta$ are parameters of the systems and $g$ is a nonlinear function of its arguments. 
As in the case of R\"ossler systems we couple the $x_1$ component.
The model system is chosen so that the synchronizd state corresponds to the fixed point $x^*=y^*=(0,0)$ for small values of $\varepsilon$ and we observe a desynchronization transition as $\varepsilon$ increases \cite{note2}. 
For this to happen the parameters of the system must obey the conditions;
$a+d < 0, \; d > 0, \;
(ad - bc) > 0$.
Under these conditions, the fixed point $(0,0)$ becomes unstable at the critical coupling constant $\varepsilon_c = \varepsilon_{c2} = \frac{1}{2}(a-\frac{bc}{d})$. 

\subsection{Quadratic nonlinearity}

We first consider quadratic nonlinearity \cite{note3},
\begin{equation}
g(u_1,u_2) = \alpha(u_1^2 + \beta u_1 u_2 + u_2^2)
\label{nonlinear_quadratic}
\end{equation}
With quadratic nonlinearity, the model system has three fixed points. One is $(0,0,0,0)$ which is also a fixed point $(0,0)$ of the uncoupled systems. 
The other two fixed points are given by
\begin{eqnarray}
(x_1^{*},x_2^{*},y_1^{*},y_2^{*}) & = & (\frac{A}{2}
\pm \sqrt{B},   
-\frac{a-\varepsilon}{b}x_1^{*} - \frac{\varepsilon}{b}y_1^{*},A - x_1^{*},
\nonumber \\ & & 
 -\frac{a-\varepsilon}{b}y_1^{*} - \frac{\varepsilon}{b}x_1^{*} )
\label{fixedpt2}
\end{eqnarray} 
where
$A = -\frac{b(2d\varepsilon-ad+bc)}{\alpha(b^2-\beta b (a-\varepsilon)+a(a-2\varepsilon))}$,
$B(\varepsilon) = \frac{A^2}{4} 
 - \frac{\varepsilon^2A^2}{W}
- \frac{bd\varepsilon A}{\alpha W}$, 
$W=a^2+b^2+4\varepsilon^2-4a\varepsilon - \beta b(a - 2\varepsilon)$.

For $\varepsilon < \varepsilon_c$, the fixed point $(0,0)$ is stable and it becomes unstable at the critical coupling constant $\varepsilon_c = \varepsilon_{c2}$. 
For $\varepsilon > \varepsilon_c$ the coupled system has two stable fixed points given by Eqs.~(\ref{fixedpt2}).

\begin{figure}[t]
\includegraphics[width = 0.96\columnwidth]{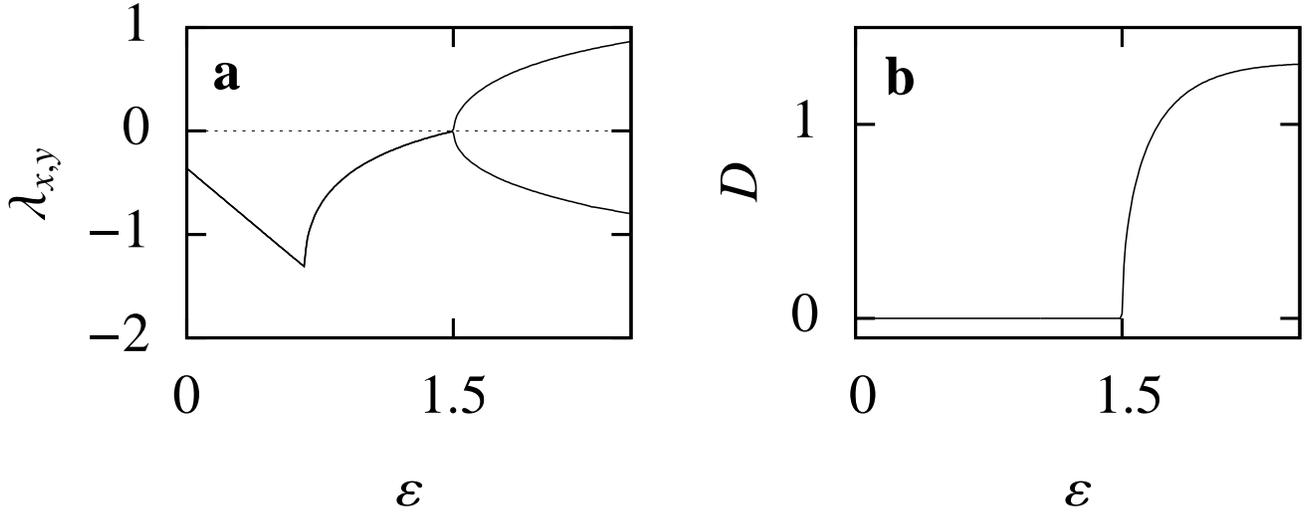}
\caption{\label{fig:five} 
{\bf a}. 
Two largest system transverse Lyapunov exponents, $\lambda_{x,y}$ of the model, Eq.~(\ref{model}) with quadratic nonlinearity, Eq.~(\ref{nonlinear_quadratic}, 
are shown as a function of the coupling parameter $\varepsilon$. 
Other parameters are $a = -1.00, b = -2.00, c = 1.00, d = 0.50, \alpha = -1.00, \beta = 2.00$. 
{\bf b}. 
The distance $D$ between the fixed points of the two systems (Eq.~(\ref{fixedpt2})) as a function of the coupling constant $\varepsilon$.}
\end{figure}

The STLE can be obtained by writing equations for 
the difference vector $z=x-y$ as in Eqs.~(\ref{stle-x}) and~(\ref{stle-y}). The largest STLEs are given by 
\begin{eqnarray}
\lambda_{x,y} \approx \pm C \sqrt{\varepsilon - \varepsilon_c} + O(\varepsilon-\varepsilon_c),
		\label{eq:sixteen}
\end{eqnarray}
where $ C = \frac{\frac{2\alpha}{b}(a^2+b^2+4\varepsilon_c^2-4a\varepsilon_c+2\beta b \varepsilon_c-\beta a d)}{(a+d-2\varepsilon_c)+\frac{\alpha(2a - \beta b)}{2b}A + \frac{\alpha}{2}(\beta b - 2 a + 4 \varepsilon_c)\sqrt{B}}  \sqrt{F},$,
$F=\frac{4b^2d^4(ad-bc)}{\alpha^2 G H}$,
$G = d(a^2+b^2-\beta a b) + (ad-bc)(\beta b - 2a)$ and $H = d^2(a^2+b^2-\beta a b) - (ad-bc)(ad+bc-2\beta b d)$.
Figure \ref{fig:five}a shows the largest STLE $\lambda_{x,y}$ as a function of the coupling constant $\varepsilon$.
For $\varepsilon < \varepsilon_c$, $\lambda_{x,y}$ are negative and equal. 
At $\varepsilon = \varepsilon_c$, they are zero and for $\varepsilon > \varepsilon_c$, one of the STLE is positive while the other is negative. 
This behavior of $\lambda_{x,y}$ is similar to that of the desynchronization transition in the coupled R\"ossler system seen in Fig.~\ref{fig:one}.

The distance between the attractors of the two systems, i.e. between $(x_1^{*}, x_2^{*})$ and $(y_1^{*}, y_2^{*})$, is given by
\begin{eqnarray}
D & = &
\frac{\sqrt{(b^2+(2\varepsilon_c-a)^2)F}}{b}\sqrt{\varepsilon - \varepsilon_c} + O(\varepsilon - \varepsilon_c)
\label{eq:eighteen}
\end{eqnarray}
Figure~\ref{fig:five}b plots the distance $D$ as a function of the coupling constant $\varepsilon$. 
Thus, for $\varepsilon > \varepsilon_c$, $D \propto \sqrt{\varepsilon - \varepsilon_c}$.

At the desynchronization bifurcation in the model system the fixed point $(0,0)$ becomes unstable and two new stable fixed points emerge. 
The distance between the stable fixed points grows proportional to $\sqrt{\varepsilon - \varepsilon_c}$. 
These are the characteristic features of the supercritical pitchfork bifurcation \cite{DDJoseph,RevModPhys.63.991}. 
This bifurcation takes place in the transverse manifold. 
This can be seen by noting that the three fixed points of the the model system, 
can also be obtained from the equation satisfied by the transverse component $z_1^*$ as
\begin{equation}
z_1^{*} (B(\varepsilon) - (z_1^*)^2) =0
\label{cubic} 
\end{equation}
This is a cubic equation and $B(\varepsilon) \propto (\varepsilon - \varepsilon_c)$ since $B(\varepsilon_c)=0$. 
This equation is exactly the normal form of a pitchfork bifurcation \cite{DDJoseph,RevModPhys.63.991}. 
Similar equation can be written for $z_2^{*}$. 

The proposed model with quadratic nonlinearity will show supercritical pitchfork bifurcation when 
$\beta >= \frac{(ad-bc)(a^2+b^2) -2\varepsilon(a^2d-b^2d-2abc)}{a(ad-bc)+ 2b c\varepsilon}$. 
Otherwise it will undergo subcritical pitchfork bifurcation.

\subsection{Cubic nonlinearity}

\begin{figure}[h]
\includegraphics[width=.9\columnwidth]{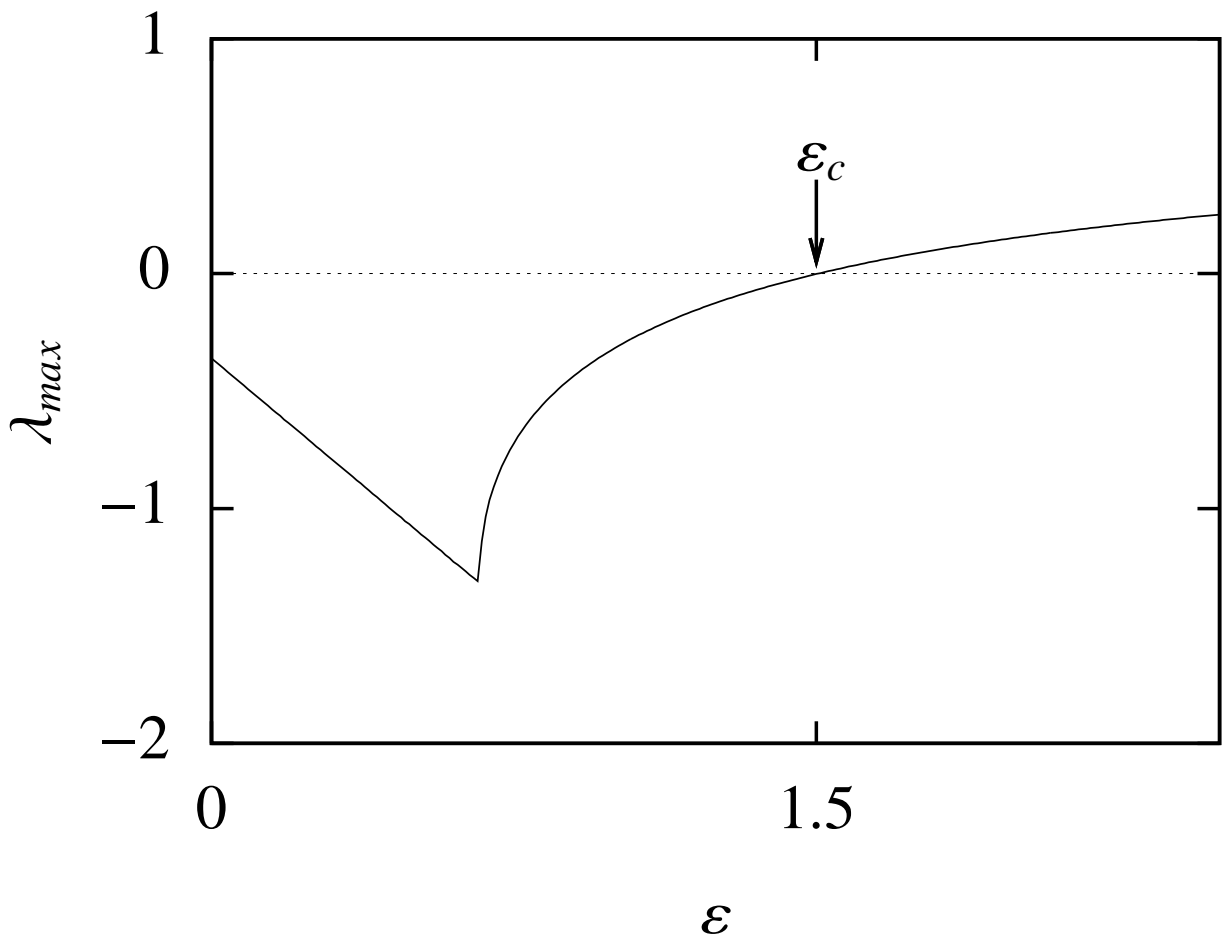}
\caption{\label{fig:model_cubic_msf} The largest transverse Lyapunov exponent, $\lambda_{max}$ is plotted with the coupling strength $\varepsilon$ 
for the model, Eq.~(\ref{model}) with cubic nonlinearity (Eq.~\ref{nonlinear_cubic}).
The critical coupling strength is  $\varepsilon_{c}=1.5$.
When $\varepsilon>\varepsilon_{c}$ the $\lambda_{max}$ is positive and synchronous state become unstable.
The system parameters are 
$a' = -1.00, b' = -2.00, c' = 1.00, d' = 0.50, \alpha = -1.00, \beta_{1} = 3.00, \beta_{2} = 3.00$. }
\end{figure}

\begin{figure}[h]
\includegraphics[width=.96\columnwidth]{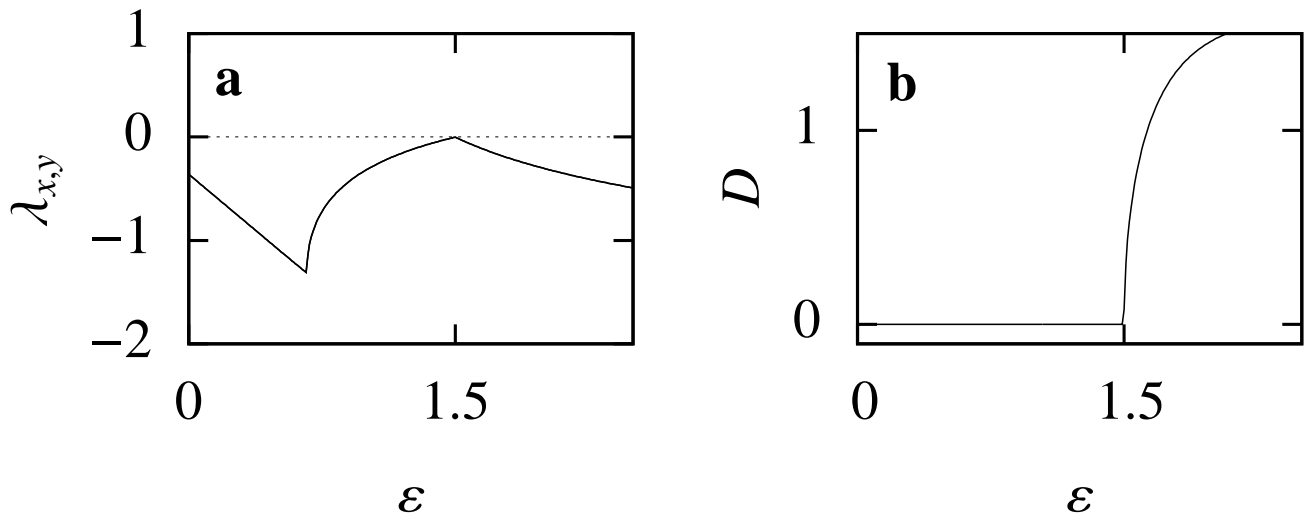}
\caption{\label{fig:model_cubic}
{\bf a}. 
Two largest system transverse Lyapunov exponents, $\lambda_{x,y}$ of the model, Eq.~(\ref{model}) with cubic nonlinearity, Eq.~(\ref{nonlinear_cubic}), 
are shown as a function of the coupling parameter $\varepsilon$. 
Other parameters are $a' = -1.00, b' = -2.00, c' = 1.00, d' = 0.50, \alpha = -1.00, \beta_{1} = 3.00, \beta_{2} = 3.00$. 
{\bf b}. 
The distance $D$ between the fixed points of the two systems for cubic nonlinearity as a function of the coupling constant $\varepsilon$.}
\end{figure}

We now consider cubic nonlinearity
\begin{equation}
g(u_1,u_2) = \alpha(u_1^3 + \beta_{1} u_1^{2} u_2 + \beta_{2} u_{1} u_{2}^{2} + x_2^3)
\label{nonlinear_cubic}
\end{equation}
In Fig.~\ref{fig:model_cubic_msf} the largest transverse 
Lyapunov exponent $(\lambda_{max})$ of this system is plotted with the coupling strength. As $\varepsilon$ crosses the critical coupling 
strength $(\varepsilon_{c})$ the largest transverse Lyapunov exponent become positive and synchronized state become unstable.
In Fig.~\ref{fig:model_cubic}a we plot the two largest systems' transverse Lyapunov exponents ($\lambda_{x}$ and $\lambda_{y}$) of the model system 
given by Eq.~\ref{model} with cubic nonlinearity as a function of the coupling strength $\varepsilon$. 
Here we can find that the exponents have same value for all coupling strengths and everywhere they are negative, except 
at the critical coupling strength, $\varepsilon_{c}$ where both of them are zero.

In the desynchronized state one can calculate the stable solutions analyticallly for the cubic nonlinearity (Eqs.~(\ref{model}) and~(\ref{nonlinear_cubic})).
The fixed points are given by,
\begin{eqnarray}
x_{1}^{*}&=& \pm \frac{b\sqrt{2d(\varepsilon-\varepsilon_{c})}}{\sqrt{F}} \nonumber\\
x_{2}^{*}&=& - \frac{(a-2\varepsilon)}{b}x_{1}^{*} \nonumber\\
y_{1}^{*}&=& -x_{1}^{*}\nonumber\\
y_{2}^{*}&=& -x_{2}^{*},
\label{eq:model_cubic_sol}
\end{eqnarray}
where $\varepsilon_{c}=\frac{1}{2}(a-\frac{bc}{d})$ and
$F=\alpha\{(a-2\varepsilon)^{3}-\beta_{2}b(a-2\varepsilon)^{2}+\beta_{1}b^{2}(a-2\varepsilon)-b^{3}\}$. 
In the synchronized state the systems synchronize in the $(0,0)$ solution. When the coupling strength $\varepsilon$ crosses the critical value $\varepsilon_{c}$ 
the systems undergo desynchronizaion bifurcation as depicted in Fig.~\ref{fig:model_cubic_msf}, 
but all STLEs are negative (Fig.~\ref{fig:model_cubic}a). 
So, the individual systems are stable. The distance between the two fixed points is proportional to $\sqrt{\varepsilon - \varepsilon_c}$ and is shown in Fig.~\ref{fig:model_cubic}b.

We can calculate the STLEs for cubic nonlinearity by considering the transverse component $z=x-y$,
\begin{eqnarray}
\lambda_{x,y} &\approx& \frac{2Fd\{F-(a-2\varepsilon)G-bH\}}{\{(a+d-2\varepsilon)F+(\varepsilon-\varepsilon_{c})G\}} 
(\varepsilon-\varepsilon_{c}) \nonumber \\
&+& {\rm higher~order} \label{eq:model_cubic_stle}
\end{eqnarray}
where, $G=3(a-2\varepsilon)^{2}+\beta_{1}b^{2}-2\beta_{2}b(a-2\varepsilon)$ and \\
$H=3b^{2}+\beta_{2}(a-2\varepsilon)^{2}-2\beta_{1}b(a-2\varepsilon)$. 
The STLEs for cubic nonlinearity~(\ref{nonlinear_cubic}) have linear dependence on the parameter after 
the desynchronization bifurcation takes place and both are negative.

\subsection{Comparison with coupled R\"ossler systems}

We now compare the results for the model system with that of two coupled R\"ossler systems.
Comparing Figures \ref{fig:four}b, ~\ref{fig:five}b and~\ref{fig:model_cubic}b, we see that for both the model and the coupled R\"ossler systems, for $\varepsilon > \varepsilon_c$, $D \propto \sqrt{\varepsilon - \varepsilon_c}$.We note that $D$ may be taken as the distance between the attractors of the two systems or the distance between the two solutions obtained by the $x\rightleftarrows y$ exchange symmetry.
For the coupled R\"ossler systems these solutions are chaotic while for the model system they are fixed points. The nature of these solutions depends on the synchronization manifold. However, the desynchronization bifurcation takes place in the transverse manifold were the both the coupled R\"ossler systems and our model show a very similar behavior. 

For the coupled R\"ossler systems we can carry out an approximate analysis. 
We write equations for the difference and sum of the variables of the two systems, $z=u^{(1)}-u^{(2)}$ and $s=u^{(1)}+u^{(2)}$, and then treat $z$ and $s$ as constants near the desynchronization bifurcation. 
This gives a cubic equation for the transverse components as $z_2 ({\cal B} -z_2^2)$ where ${\cal B}$ depends on the parameters. 
The condition ${\cal B}=0$ gives $\varepsilon_{c2} \sim 3.33\ldots$ which is somewhat larger than the observed value of 3.002 of the desynchronization bifurcation.

Thus both the transitions in the coupled R\"ossler systems and our model can be identified as supercritical pitch-fork bifurcations of the transverse manifold.

The nature of the nonlinearity can be identified using STLEs defined by us.
Comparing the behavior of STLEs for $\varepsilon > \varepsilon_c$ in Figs.~\ref{fig:one}, \ref{fig:five}a and~\ref{fig:model_cubic}a,
we see that the behavior of SLTEs for the coupled R\"ossler systems matches with that of our model with quadratic nonlinearity, but not with the cubic nonlinearity.

We find the the form used in Eq.~(\ref{model}) with  quadratic~(Eq.~(\ref{nonlinear_quadratic})) or cubic~(Eq.~(\ref{nonlinear_cubic})) nonlinearity, is the simplest form we could get for the desynchronization bifurcation of the transverse manifold. The model also 
gives the standard normal form (Eq.~(\ref{cubic})), of the pitchfork bifurcation for the transverse component. 
Hence, the model may be treated as the normal form for the desynchronization bifurcation \cite{note6}. 
We note that the coupled R\"ossler systems and the model with quadratic nonlinearity have similar properties. 
Hence, we conjecture that our model of Eq.~(\ref{model}) with quadratic nonlinearity~(Eq.(\ref{nonlinear_quadratic})) is the normal form for the desynchronization bifurcation of the coupled R\"ossler systems.

\section{\label{discussion}DISCUSSION}

From the discussion above, we conclude that the desynchronization bifurcation of the coupled model system, 
Eq.~(\ref{model}) as well as the coupled R\"ossler systems, Eq.~(\ref{eq:seven}), are supercritical pitchfork bifurcations of the transverse manifold. 
The synchronization manifold decides the nature of the attractor which is chaotic for 
the coupled R\"ossler systems while it is periodic (fixed points) for our model system. 

We have presented the analysis for symmetric coupling with $\varepsilon = \varepsilon_1 = \varepsilon_2$. 
If instead we take asymmetric coupling $\varepsilon_1 \neq \varepsilon_2$, 
the nature of the desynchronisation bifurcation does not change. 
This is because this bifurcation takes place in the transverse manifold defined by 
the difference vector $z$ and in the equation for $z$, (Eq.~(\ref{eq:four})), we only have the sum $\varepsilon_1 + \varepsilon_2$. 
We also note that for $\varepsilon_1 \neq \varepsilon_2$, the $x\Leftrightarrow y$ 
exchange symmetry exists in the trasverse component though not in the longitudinal component.

We find the the form used in Eq.~(\ref{model}) is the simplest form we could get for the desynchronization bifurcation and also, 
we get the standard normal form (Eq.~(\ref{cubic})), of the pitchfork bifurcation for the transverse component. 
Hence, the model may be treated as the normal form for the desynchronization bifurcation. 
We can further simplify the model by choosing $a=-1, \, c=1,\, \alpha = \pm 1$. 
We note that the coupled R\"ossler systems and the model have similar properties. 
Hence, we conjecture that Eq.~(\ref{model}) with the quadratic nonlinearirty (Eq.(\ref{nonlinear_quadratic})) is the normal form for the desynchronization bifurcation of the coupled R\"ossler systems.

Let us now consider coupled R\"ossler systems on a network. 
Consider $n$ coupled R\"{o}ssler oscillators. Denoting the variables by $u^{(j)}, \, j=1,2,\ldots,N$, the equations can be written as 
\begin{eqnarray}
\dot{u}^{(j)} = f(u^{(j)}) + \varepsilon \sum_{k} J_{jk} \Gamma(u^{(k)} - u^{(j)}), \label{eq:nineteen}
\end{eqnarray}
where $J$ is the coupling matrix.
The analysis of Pecora and Carrol \cite{PhysRevLett.80.2109} shows that the equations for 
the transverse manifold can be cast into a general form of a master equation and is the same as that for the two coupled systems. 
Thus, the present analysis should be applicable for the desynchronization transition for coupled systems on a network. 
How do the attractors of the different systems split for $\varepsilon > \varepsilon_{c2}$? 
Consider three mutually coupled R\"ossler systems.
We observe an interesting phenomena of symmetry breaking. 
In this case at the desynchronization bifurcation we still get splitting of the attractors into two as in Fig.~\ref{fig:four}a, 
with two oscillators on one side and the remaining oscillator on the other side. 
The two oscillators on the same side remain synchronized \cite{note1}.
We find that the distance between the center of these oscillators varies with the coupling in the same fashion as in Eq.~(\ref{eq:eight}). 
When four oscillators are coupled in a reactangle then this desynchronization bifurcation takes place between two pairs of oscillators. 
The oscillators in the same pair remain synchronized.

\section{\label{conlcusion}conclusion:}

To conclude, we have analysed the desynchronization bifurcation in the coupled R\"{o}ssler systems. 
We give a simple model of coupled integrable systems which shows a similar phenomena. 
The model may be treated as the normal form for the desynchronization bifurcation. 
After the desynchronization bifurcation the attractors of the coupled systems split into two and start moving away from each other. 
We define system transverse Lyapunov exponents corresponding to the difference vector of the variables of the systems. 
For $\varepsilon > \varepsilon_c$ and quadratic nonlinearity, the STLE for one system becomes positive while that for the other system becomes negative. While for $\varepsilon > \varepsilon_c$ and cubic nonlinearity, the STLEs of both systems are negative.
From the analysis of the distance between the two attractors which is proportional to $\sqrt{\varepsilon - \varepsilon_c}$, the behavior of SLTEs
and the cubic form for the transverse components, we conclude that the desynchronization bifurcation in 
the coupled R\"ossler systems is a pitchfork bifurcation of the transverse manifold and has the normal form of our model with quadratic nonlinearity.

\noindent
\section*{\label{acknowledgements}Acknowledgements}
The authors thank Christophe Letellier for useful discussions.


\begin{thebibliography}{99}
\bibitem{note3}We note that the we have the most general form of the quadratic nonlinear since the coefficients of $x_1^2$ and $x_2^2$ can be made equal by a suitable scaling.
\bibitem{note6}We can further simplify the model by choosing $a=-1, \, c=1,\, \alpha = \pm 1$.
\end{thebibliography}
\end{document}